\documentstyle[12pt]{article}
\newcommand{\beq}{\begin{equation}}
\newcommand{\eeq}{\end{equation}}
\newcommand{\bea}{\begin{eqnarray}}
\newcommand{\eea}{\end{eqnarray}}
\newcommand{\bay}{\begin{array}}
\newcommand{\eay}{\end{array}}
\begin{document}
\begin{titlepage}
\begin{flushleft}
MZ-TH/94-16\\
May 1994
\end{flushleft}
\begin{center}
\large
\bf
On the $1/m^2$ corrections to the form-factors
in $\Lambda_b\to\Lambda_ce\bar\nu_e$ decays\\[2cm]
\rm
J.G.K\"orner$^1$ and D.Pirjol$^2$\\[.5cm]
Johannes Gutenberg-Universit\"at\\
Institut f\"ur Physik (THEP), Staudingerweg 7\\
D-55099 Mainz, Germany\\[2cm]
\normalsize
\bf
Abstract\\[0.5cm]
\rm
\small
\end{center}
Model-independent bounds are presented for the form-factors describing the
semileptonic decay $\Lambda_b\to\Lambda_ce\bar\nu_e$ at the end-point of
the lepton energy spectrum. For the axial-vector form-factor $f_1^A(1)$ we
obtain $f_1^A(1)\leq 0.960\div 0.935$, where the uncertainty arises from the
value of $\mu_\pi^2$, the average kinetic energy of the $b$ quark in
the $\Lambda_b$ baryon. A bound is given on the forward matrix element
of the axial current in a $\Lambda_b$ baryon which is relevant for the
inclusive semileptonic decays of the polarized $\Lambda_b$ decays.\\[4cm]
\footnotesize
$^1\,$Supported in part by the BMFT, Germany under contract 06MZ730\\
$^2\,$Supported by the Graduiertenkolleg Teilchenphysik, Universit\"at Mainz\\
\normalsize
\end{titlepage}

 {\bf 1.} Recently it has been shown \cite{1} how to obtain model-independent
bounds on form-factors describing semileptonic transitions of heavy hadrons.
It had been known for some time that some of the form-factors
describing the decay $B\to D^*e\bar\nu_e$ do not receive $1/m$ corrections
at the so-called ``no-recoil'' kinematical point \cite{Luke}. The only
corrections to
the heavy-quark symmetry limit appear at order $1/m^2$ and they have been
evaluated in the past by making use of model-dependent results
\cite{FN,HS1,HS2,HS3}.
In the paper \cite{1} a model-independent bound on the magnitude of these
corrections has been given. The aim of this paper is to present a similar
bound on the $1/m^2$ corrections to the heavy quark symmetry predictions
for the semileptonic decay $\Lambda_b\to\Lambda_c e\bar\nu_e$.\\[0.5cm]

   {\bf 2.} We begin by describing the kinematics of the decay. The
$\Lambda_b$ baryon is polarized with the spin vector $s$  and is taken to
be at rest in
the laboratory frame. It decays into a final hadronic state containing one
charmed quark of invariant mass $m_X$ and a lepton pair $e\bar\nu_e$ of
4-momentum $q$. The inclusive decay width summed over all final hadronic
states with the  same invariant mass can be expressed in terms of the
hadronic tensor
\bea
W_{\mu\nu} = (2\pi)^3\sum_X\langle\Lambda_b(v,s)|J_\mu^\dagger(0)|X\rangle
   \langle X|J_\nu(0)|\Lambda_b(v,s)\rangle\delta(p_X-p+q)
\eea
with $p_X^2=m_X^2$ and $p=m_{\Lambda_b}v$ the momentum of the $\Lambda_b$
and $J$ is the weak current. We normalize the $|\Lambda_b\rangle$ state
as usual to $v_0$. The most general decomposition into covariants for the
hadronic tensor $W_{\mu\nu}$ has the form
\bea
W_{\mu\nu} &=& -g_{\mu\nu}W_1 + v_\mu v_\nu W_2 -
i\epsilon_{\mu\nu\alpha\beta}v^\alpha q^\beta W_3 + q_\mu q_\nu W_4 +
(q_\mu v_\nu+q_\nu v_\mu) W_5\nonumber\\
&-&q\cdot s\left[-g_{\mu\nu}G_1 + v_\mu v_\nu G_2 -
i\epsilon_{\mu\nu\alpha\beta}v^\alpha q^\beta G_3 + q_\mu q_\nu G_4 +
(q_\mu v_\nu+q_\nu v_\mu) G_5\right]\\
&+&(s_\mu v_\nu+s_\nu v_\mu) G_6
+ (s_\mu q_\nu+s_\nu q_\mu) G_7
+i\epsilon_{\mu\nu\alpha\beta}v^\alpha s^\beta G_8
\nonumber\\
&+&i\epsilon_{\mu\nu\alpha\beta}q^\alpha s^\beta G_9\,.\nonumber
\eea
Here the same parametrization has been used as in \cite{2}.

  The hadronic tensor $W_{\mu\nu}$ is given by the discontinuity of the
forward scattering amplitude
\bea
T_{\mu\nu}(q\cdot v, q^2) = -i\int\mbox{d}^4\! xe^{-iq\cdot x}
\langle\Lambda_b(v,s)|TJ_\mu^\dagger(x) J_\nu(0)|\Lambda_b(v,s)\rangle\,,
\eea
across the cut in the complex plane of the variable $v\cdot q$
which corresponds to the process under consideration. In our case the
cut extends from $\sqrt{q^2}$ to $1/(2m_{\Lambda_b})(m_{\Lambda_b}^2-
m_{\Lambda_c}^2+q^2)$. The invariant mass of the final hadronic state
varies from $m_{\Lambda_c}$ on the right edge of the cut up to
$m_{\Lambda_b}-\sqrt{q^2}$ on the left edge.

  We will parametrize the forward scattering amplitude $T_{\mu\nu}$
in terms of invariant form-factors $T_{1,5}$ and $S_{1,9}$ (defined in
analogy to $W_{1,5}$ and $G_{1,9}$ in (2)).

  For our purposes it will prove more convenient to study the analyticity
properties of $T_{\mu\nu}$ as a function of complex $v\cdot q$ at fixed
velocity $v'$ of the final hadronic state (instead of $q^2$). In these
variables,
for a given $\omega=v\cdot v'$, the cut corresponding to the physical decay
process ranges between
\beq
m_{\Lambda_b}\frac{\sqrt{\omega^2-1}}{\omega+\sqrt{\omega^2-1}} \leq
v\cdot q \leq m_{\Lambda_b}-m_{\Lambda_c}\omega\,.
\eeq
The invariant mass of the final hadronic state correspondingly takes
values between
\beq
m_{\Lambda_b}\frac{1}{\omega+\sqrt{\omega^2-1}} \geq m_X \geq m_{\Lambda_c}
\eeq
and $q^2$ is a function of $v\cdot q$ along the cut, given by
\beq
q^2= -m_{\Lambda_b}(m_{\Lambda_b}-2v\cdot q)+\frac{1}{\omega^2}
(m_{\Lambda_b}-v\cdot q)^2\,.
\eeq

  The discontinuity of $T_{\mu\nu}$ across this cut is related to the
hadronic tensor (1) by
\beq
W_{\mu\nu} = -\frac{1}{2\pi i}\mbox{disc}\,T_{\mu\nu}\,.
\eeq

  In the following we will construct two sum rules for the form-factors
describing the
semileptonic decays of the $\Lambda_b$ baryon in analogy to the ones
presented in \cite{1}. The method is based on the positivity of certain
combinations of invariant form-factors in (2). These can be obtained from
inequalities of the form
\beq
n^\mu n^{*\nu} W_{\mu\nu} \geq 0
\eeq
and follow from the fact that this quantity is directly related to
the decay rate $\Lambda_b\to W(q,n_\mu)X_c$
of the $\Lambda_b$ baryon into a virtual W boson with the polarization
vector $n$. Let us work in the rest frame of the $\Lambda_b$ baryon.

  One positive-definite combination of invariant form-factors in (2) is
obtained by using in (8) a longitudinal polarization vector.
This gives
\bea
W_L = W_1+\frac{q_0^2-q^2}{q^2}W_2 - q\cdot s\left(
 G_1+\frac{q_0^2-q^2}{q^2}G_2\right) + 2\frac{q_0}{q^2}(s\cdot q)G_6\,.
\eea

  Another possible choice is the total decay rate into a circularly
polarized W boson, which is proportional to
\bea
W_{T_{L,R}} = W_1 - q\cdot sG_1 \pm \left( \sqrt{q_0^2-q^2}(W_3 -
q\cdot sG_3) + \frac{q\cdot s}{\sqrt{q_0^2-q^2}}(G_8+q_0G_9)\right)\,
{}.
\eea
The total decay rate into a transversally polarized W is
\bea
W_T = 2W_1 - 2q\cdot s G_1\,.
\eea

  Finally, the case of a scalar (or ``temporal'') polarization gives
\bea
W_0 &=& -W_1+\frac{q_0^2}{q^2}W_2 + q^2W_4 + 2q_0W_5\\
 &-& q\cdot s\left(-G_1+\frac{q_0^2}{q^2}G_2 + q^2G_4 + 2q_0G_5
\right) + 2\frac{q_0}{q^2}(s\cdot q)G_6 + 2(s\cdot q)G_7\,.\nonumber
\eea

  These quantities must be positive for any possible orientation of the
spin vector $\vec s$ with respect to $\vec q$ and for any current $J$ in
(1).\\[0.5cm]

   {\bf 3.} The forward scattering amplitude $T_{\mu\nu}$ can be given a
representation in terms of physical intermediate states as
\bea
T_{\mu\nu} &=& \sum_{X(p_X=m_{\Lambda_b}v-q)}
\frac{\langle\Lambda_b|J_\mu^\dagger(0)|X\rangle
      \langle X|J_\nu(0)|\Lambda_b\rangle}{m_{\Lambda_b}-E_X-q_0+i\epsilon}
\\
&+& \sum_{X(p_X=m_{\Lambda_b}v+q)}
\frac{\langle\Lambda_b|J_\nu(0)|X\rangle
      \langle X|J_\mu^\dagger(0)|\Lambda_b\rangle}
   {m_{\Lambda_b}-E_X+q_0+i\epsilon}\,.\nonumber
\eea
Only the first term above gives a contribution to the discontinuity across
the cut corresponding to the decay process of interest. We will be mainly
interested in the contribution of the intermediate state $\Lambda_c$.
   We will consider first
the case of an axial current $J_\mu=\bar c\gamma_\mu\gamma_5b$.
The corresponding matrix element of the axial current in (13) can be
parametrized as usual in terms of 3 form-factors, defined through
\bea
\langle\Lambda_c(v',s')|\bar c\gamma_\mu\gamma_5 b|\Lambda_b(v,s)\rangle
=\bar u(v',s')[f_1^A\gamma_\mu + f_2^Av_\mu + f_3^A v'_\mu ]\gamma_5u(v,s)\,.
\eea
We have used a parametrization appropriate to the heavy-mass limit, in
terms of the velocities of the particles involved. The spinors are
normalized so that $\bar uu=1$.

  By taking the imaginary part of (13), the hadronic tensor $W_{\mu\nu}$ can
be also written as a sum over intermediate states. The contribution of the
lowest-lying state $\Lambda_c$ to the positive-definite combinations of
structure functions defined in (9,10,12) is
\bea
W_L &=& \frac{\omega+1}{2q^2}\left((m_{\Lambda_b}-m_{\Lambda_c})f_1^A
-m_{\Lambda_c}(\omega-1)f_2^A-m_{\Lambda_b}(\omega-1)f_3^A\right)^2\\
W_{T_{L,R}} &=& |f_1^A|^2\frac{1+\omega}{2}(1\pm\hat n\cdot \vec s\,)\\
W_0 &=& \frac{\omega-1}{2q^2}\left((m_{\Lambda_b}+m_{\Lambda_c})f_1^A
-(m_{\Lambda_b}-m_{\Lambda_c}\omega)f_2^A-(m_{\Lambda_b}\omega-
m_{\Lambda_c})f_3^A\right)^2\,.
\eea
A factor of $1/\omega\delta(q_0-m_{\Lambda_b}+m_{\Lambda_c}\omega)$ has been
removed from these expressions.
Here $\hat n$ is a unit vector collinear with the vector $\vec q$. If
$|\vec q\,|=0$, it defines the quantization axis along which the spin
of the virtual W boson is aligned. The above formulas can be simply related
to helicity amplitudes for the process $\Lambda_b(\lambda_{\Lambda_b})\to
\Lambda_c(\lambda_{\Lambda_c})+W(\lambda_W)$ listed for different other
cases of physical interest in \cite{BKKZ}.

  For the case of a vector current $J_\mu=\bar c\gamma_\mu b$, we define
in an analogous way to (14)
\bea
\langle\Lambda_c(v',s')|\bar c\gamma_\mu b|\Lambda_b(v,s)\rangle
=\bar u(v',s')[f_1^V\gamma_\mu + f_2^Vv_\mu + f_3^V v'_\mu ]u(v,s)\,.
\eea
In terms of these form-factors, the contribution of the intermediate state
$\Lambda_c$ reads
\bea
W_L &=& \frac{\omega-1}{2q^2}\left((m_{\Lambda_b}+m_{\Lambda_c})f_1^V
+m_{\Lambda_c}(\omega+1)f_2^V+m_{\Lambda_b}(\omega+1)f_3^V\right)^2\\
W_{T_{L,R}} &=& |f_1^V|^2\frac{\omega-1}{2}(1\pm\hat n\cdot \vec s)\\
W_0 &=& \frac{\omega+1}{2q^2}\left((m_{\Lambda_b}-m_{\Lambda_c})f_1^V
+(m_{\Lambda_b}-m_{\Lambda_c}\omega)f_2^V+(m_{\Lambda_b}\omega-
m_{\Lambda_c})f_3^V\right)^2\,.
\eea
As before, a factor of $1/\omega\delta(q_0-m_{\Lambda_b}+m_{\Lambda_c}\omega)$
has to be added on the r.h.s..\\[0.5cm]

   {\bf 4.} On the other hand, it has been recently shown
\cite{CGG,6,3,2,mannel} that it is
possible to reliably calculate the forward scattering matrix element (3) in
a region which is far away from the physical cuts. The result takes the
form of an expansion in powers of $1/m_{c,b}$ and depends on a few
nonperturbative matrix elements of dimension-6 operators taken between
heavy hadron states. For the case of an axial (vector) current
$J_\mu=\bar c\gamma_\mu\gamma_5 b$, the
results for $T_{1,5}$ can be taken from \cite{3}.
As for the spin-dependent structure functions one can see from (9,10,12)
that at the equal-velocity point $\omega=1$, only $G_{8,9}$
contribute. The corresponding amplitudes $S_{8,9}$ can be straightforwardly
calculated following the method given in \cite{3,2} with the result
(for $J_\mu=\bar c\gamma_\mu\gamma_5b$; the result for the vector current
case can be obtained by making the replacement $m_c\to -m_c$)
\bea
S_8 &=& \frac{1}{\Delta}\left(1+\frac{\mu_s^2}{m_b^2}+\frac{\mu_\pi^2}
{6m_b^2}\right)(m_b+m_c) + \frac{5\mu_\pi^2}{3m_b\Delta^2}v\cdot q(m_b+m_c)
\nonumber\\
&+&\frac{4\mu_\pi^2}{3\Delta^3}[(v\cdot q)^2-q^2](m_b+m_c)\\
S_9 &=&-\frac{1}{\Delta}\left(1+\frac{\mu_s^2}{m_b^2}\right) -
\frac{\mu_\pi^2}{m_b\Delta^2}\left(v\cdot q+\frac23(m_b+m_c)\right)
\nonumber\\
&-&\frac{4\mu_\pi^2}{3\Delta^3}[(v\cdot q)^2-q^2]
\eea
where
\bea
\Delta = m_b^2-2m_bq_0+q^2-m_c^2+i\epsilon
\eea
and
\bea
\mu_\pi^2 = -\langle\Lambda_b(v,s)|\bar h_b(iD)^2
h_b|\Lambda_b(v,s)\rangle_{spin-averaged}\,.
\eea
$\mu_\pi^2$ represents the average kinetic energy of the $b$ quark inside a
$\Lambda_b$ baryon. At present its numerical value is only poorly known.
It has been shown, on quite general grounds, that it must be positive
\cite{BSUV}.
A previous estimate \cite{FN} of the corrections of order $1/m^2$ to the
$\Lambda_b\to\Lambda_c$ form-factors used $\mu_\pi^2\simeq -1$ GeV$^2$.
The corresponding parameter in the $B-B^*$ system has the approximate
value 0.5 GeV$^2$. Using the mass formula
\beq
m_{\Lambda_b} = m_b+\bar\Lambda + \frac{\mu_\pi^2}{2m_b}
\eeq
and a similar one for the $\Lambda_c$ baryon, together with the mass values
\cite{DPG} $m_{\Lambda_b}=5.641$ GeV, $m_{\Lambda_c}=2.285$ GeV, $m_b=4.8$
GeV and $m_c=1.39$ GeV, yields
\beq
\mu_\pi^2 = 0.211\,\mbox{GeV}^2\,.
\eeq
For the purpose of illustration we will vary the value of $\mu_\pi^2$ between
0.211 and 0.500 GeV$^2$.

  The other parameter in (22,23) is $\mu_s^2$, which is defined as the
$1/m_b^2$-correction to the forward matrix element of the axial current in a
$\Lambda_b$ state
\beq
\langle\Lambda_b(v,s)|\bar b\gamma_\mu\gamma_5b|\Lambda_b(v,s)\rangle
=\left(1+\frac{\mu_s^2}{m_b^2}+\cdots\right)\,\bar u\gamma_\mu\gamma_5u\,.
\eeq
This matrix element does not receive any corrections at order $1/m_b$
\cite{GGW}. $\mu_s^2$ is related to the parameter $\epsilon_b$ defined in
\cite{2}
by $\mu_s^2 = m_b^2\epsilon_b$. Very little is known about its precise
numerical value. We will show below that it must be negative and will give
an upper bound for its value.

  Inserting $q^2$ from (6) in the theoretical expression for $T_{\mu\nu}$ and
taking the imaginary part yields the QCD prediction for the hadronic
tensor, which is a singular function of $q_0$ consisting of $\delta$
functions and its derivatives. Integrating the QCD prediction for $W_1$
over the physical cut in the $q_0$ complex plane at fixed $\omega=1$ one
obtains
\bea
\int\mbox{d}q_0W_1(q_0,\omega=1) =
1 - \frac{\mu_\pi^2}{6m_cm_b} - \frac{\mu_\pi^2}{4m_b^2}
- \frac{\mu_\pi^2}{4m_c^2}\,.
\eea

  As one can see from (11), this quantity is positive definite, which means
that it represents an upper bound for the contribution of any physical
state to the hadronic tensor. In particular, from (15,16) one obtains a bound
on the invariant form-factor $f_1^A(1)$ which is relevant for the decay
rate of the process $\Lambda_b\to\Lambda_c e\bar\nu_e$ near the endpoint
of the electron spectrum:
\beq
|f_1^A(1)|^2 \leq 1 - \frac{\mu_\pi^2}{6m_cm_b} - \frac{\mu_\pi^2}{4m_b^2}
- \frac{\mu_\pi^2}{4m_c^2}\,.
\eeq
This bound is similar to the bound recently derived in \cite{1}.
Numerically, this bound is
\beq
1-|f_1^A(1)|^2 \geq 0.165\mu_\pi^2\,(\mbox{GeV}^2) = (3.5\div 8.3)\%\,.
\eeq
This result is comparable with the estimate of the same quantity in
\cite{FN} which gave about 6.2\%.

   This bound can be improved if the contribution of the spin-dependent
structure functions is taken into account. Let us consider for example
the combination $W_{T_L}$ defined in (10), at $\omega=1$. The $\Lambda_c$
contribution to its integral along the cut is
\beq
\int\mbox{d}q_0W_{T_L}(q_0,\omega=1) = |f_1^A(1)|^2(1+\hat n\cdot \vec s\,)
\eeq
whereas the QCD prediction for the same quantity is
\bea
\int\mbox{d}q_0W_{T_L}(q_0,\omega=1) &=& \left(1 - \frac{\mu_\pi^2}{6m_cm_b} -
\frac{\mu_\pi^2}{4m_b^2} - \frac{\mu_\pi^2}{4m_c^2}\right)
(1+\hat n\cdot \vec s\,)\nonumber\\
&+&(n\cdot \vec s\,)\frac{1}{m_b^2}(\mu_s^2+\frac{\mu_\pi^2}{3})\,.
\eea
Requiring the positivity of this quantity for any value of $\hat n\cdot\vec
s$ gives the inequality
\bea
\mu_s^2+\frac{\mu_\pi^2}{3} \leq 0\,,
\eea
from which one obtains an upper bound on $\mu_s^2$
\beq
\mu_s^2 \leq -\frac{\mu_\pi^2}{3} \simeq (-0.070 \div -0.167)\,\mbox{GeV}^2\,.
\eeq
In \cite{FN} $\mu_s^2$ has been estimated to be $\mu_s^2=-\mu_\pi^2/3$,
 assuming that the contribution of terms arising from
double insertions of chromomagnetic operator can be neglected. The present
method gives also the sign of the corrections to this estimate.

  The inequality (34) implies that the bound (30) on $|f_1^A(1)|$ can be
improved, provided that a determination (calculation) of $\mu_s^2$ becomes
available. For the moment, we will restrict ourselves to the simple bound
(30).

   In a completely analogous way one can obtain a different bound on the
vector-current form-factors $f_{1,2,3}^V$ defined in (18). This time the
bound results from requiring the positivity of the scalar combination
of form-factors (12). From (21), the contribution of the $\Lambda_c$ state
reads
\bea
\int\mbox{d}q_0W_0(q_0,\omega=1) = (f_1^V(1)+f_2^V(1)+f_3^V(1))^2\,,
\eea
whereas the QCD analysis yields for the same quantity
\bea
\int\mbox{d}q_0W_0(q_0,\omega=1) = 1-\frac{\mu_\pi^2}{4}\left(\frac{1}{m_c}
-\frac{1}{m_b}\right)^2\,.
\eea
It has been shown \cite{GGW} that the combination of form-factors appearing
in (36) receives no $1/m$ corrections at $\omega=1$. Putting together (36)
and (37) gives a lower bound on the magnitude of the $1/m^2$ corrections to
the leading-order result:
\bea
1-(f_1^V(1)+f_2^V(1)+f_3^V(1))^2 \geq \frac{\mu_\pi^2}{4}\left(\frac{1}{m_c}
-\frac{1}{m_b}\right)^2 = 0.065\mu_\pi^2 \simeq (1.4\div 3.2)\%\,.
\eea
which is in agreement with the 19\% estimate of this quantity in \cite{FN}.
On the other hand, in \cite{HS3} $\sum f_i^V(1)$ was found to be larger than
unity. As will be shown below, the radiative corrections could account for
such an increase.

   These bounds get changed when radiative corrections are incorporated.
Their effect is to modify the heavy quark prediction for the form-factors
at leading order in $1/m$ by a multiplicative factor. To one-loop order, the
factor multiplying $f_1^A(1)$ is \cite{radiative}
\beq
\eta_A^{pert} = 1+\frac{\alpha_s}{\pi}\left(\frac{m_b+m_c}{m_b-m_c}
\log\frac{m_b}{m_c}-\frac{8}{3}\right)
\simeq 0.958\,,
\eeq
and the factor multiplying the form-factor combination $f_1^V(1)+f_2^V(1)+
f_3^V(1)$
\beq
\eta_V^{pert} = 1+\frac{\alpha_s}{\pi}\left(\frac{m_b+m_c}{m_b-m_c}
\log\frac{m_b}{m_c}-2\right)
\simeq 1.025\,
\eeq
where $\Lambda_{QCD}=0.250$ GeV has been used.
As a result we finally obtain
\bea
& &f_1^A(1) \leq 0.960 \div 0.935\\
& &f_1^V(1)+f_2^V(1)+f_3^V(1) \leq 1.005 \div 0.996\,.
\eea
In principle these bounds could be improved if the contributions of any
other states to the integrals of the positive-definite quantities (11,12)
are known or if they can be estimated in some way. In particular, the
contribution of
two-body intermediate states like $\Lambda_c\eta$ and $\Sigma_c\pi$ can be
estimated by making use of the Heavy Hadron Chiral Perturbation Theory
\cite{cho,review}.
Unfortunately, the coupling of the light Goldstone bosons to the
antitriplet of heavy baryons containing one heavy quark is at present
unknown and any estimate of these contributions will necessarily entail
model-dependent assumptions.

\end{document}